\documentstyle[floats,multicol,aps,epsf,prb]{revtex}

\begin{document}
\draft
\twocolumn[\hsize\textwidth\columnwidth\hsize\csname @twocolumnfalse\endcsname
\title{
Bosonization in the two-channel Kondo model
}
\author{A. J. Schofield}
\address{Department of Physics, Rutgers University,
Piscataway, P.O. Box 849, NJ--08855--0849}
\date{\today}
\maketitle
\begin{abstract}
The bosonization of the $S=1/2$ anisotropic
two-channel Kondo model is shown to yield two equivalent
representations of the original problem. In a straight forward
extension of the Emery-Kivelson approach, the interacting resonant
level model previously derived by the Anderson-Yuval technique is
obtained.  In addition, however, a ``($\sigma$,$\tau$)'' description
is also found. The strong coupling fixed point of the
($\sigma$,$\tau$) model was originally postulated to be related to the
intermediate coupling fixed point of the two-channel Kondo model. The
equivalence of the $\sigma$,$\tau$ model to the two-channel Kondo
model is formally established. A summary of what one may learn from
a simple study of these different representations is also given.
\end{abstract}
]

The physics of the isotropic two-channel Kondo problem has been the
focus of intense theoretical investigation since it was first known to
have an intermediate coupling fixed point\cite{noziere1}.  The
essential physics was understood at that time to be due to an
overscreening of the impurity spin leaving a residual spin object to
interact with the conduction electrons even in the strong coupling
limit. This has since been clarified by the exact solution of the
problem using the Bethe Ansatz\cite{andrei1,tsvelik1}.

Considerable insight into this exact solution was recently provided by
two apparently different approaches which both enable much of the
physics of the problem to be extracted by elementary methods. In the
first of these, Emery and Kivelson\cite{emery} found that a
bosonized version of the model could be reduced to a resonant level
model which becomes diagonal for a special value of perpendicular
coupling $J_z$.  The non-Fermi liquid thermodynamics and the Wilson
ratio of $8/3$ found near this point\cite{sengupta} are 
consistent with the exact solution.
In the second approach, Coleman {\it
et. al.}\cite{pedest} postulated a ``$\sigma+\tau$'' model, where an
$S=1/2$ impurity interacts with both the spin ($\sigma$) and charge
($\tau$) degrees of freedom of a {\em single} channel of conduction
electrons. This model has a strong--coupling fixed point which, the
authors argue, is equivalent to the intermediate coupling fixed point
of the symmetric two-channel Kondo problem. A strong coupling
expansion about this fixed point also gives a good account of the low
energy physics of the exact solution and again reproduces the correct
Wilson ratio.

In analyzing the more general case where channel anisotropy in the
couplings is allowed, a similar pattern of events has transpired.
Andrei and Jerez\cite{jerez} have obtained a solution for
the thermodynamics of this model using the full machinery of the Bethe
Ansatz. At the same time, an equivalent interacting resonant level
model was obtained by Fabrizio {\it et. al.}\cite{fabrizio} 
by the Anderson-Yuval technique and analyzed in the vicinity of 
the Toulouse point. The
($\sigma$,$\tau$) model was also extended by Coleman and
Schofield\cite{schofield95} to examine the fixed point of that model away
from the symmetric point by a duality transformation. Controversially,
the authors argued that a Fermi liquid description of the strong
coupling fixed point is not possible because of the Majorana
structure of the Anderson model they derive at strong coupling. Up
until now, the weak point in applying their argument to the
two-channel Kondo model lies in the absence of the
precise link between this model and the ($\sigma$,$\tau$) model which
they treat. The need for such a connection has become all the more pressing
following recent papers by Zhang and Hewson~\cite{zhang96} which argue
that the models have different fixed points on the basis of a comparison
of the resistivities of the two models.

In this paper, I formally derive the ($\sigma,\tau$) description from
the two-channel Kondo model using bosonization. The result is valid
for both the channel symmetric and anisotropic cases.  The 
resonant level model obtained by Fabrizio {\it et. al. }\cite{fabrizio}
is also derived by this method (a result
undoubtedly known by the authors of that paper). This gives a
foundation for understanding how these simple pictures relate to the
exact solution. In particular the explicit mapping between the 
($\sigma,\tau$) model and the two-channel Kondo model shows why the 
``resistivity'' calculated by Zhang and Hewson~\cite{zhang96} for the
($\sigma,\tau$) model has no correspondence with the two channel Kondo
model.
Recent work by Kotliar and
Si\cite{kotliar} has emphasized the importance of ``Klein'' factors in
finding Toulouse points so, for completeness, I give a precise account
of the bosonization technique including these factors.

The two-channel Kondo problem describes how an impurity
spin (taken here to be $S=1/2$) at the origin interacts with a
conduction sea of spin--half electrons of two flavors (channels). Typically one
restricts the interaction to being in the $s$-wave sector for each
flavor reducing
the problem to finding the radial wave functions of the conduction
sea. By mapping the incoming partial waves onto the positive $x$ axis
and out-going waves onto the negative $x$ axis one can
treat the problem as left-moving 1D chiral electrons interacting with an
impurity at the origin\cite{affleck91}
\begin{equation}
{\cal H}={\cal H}_0+{\cal H}_\perp + {\cal H}_\parallel +{\cal H}_B\;
,
\label{ham}
\end{equation}
where
\begin{eqnarray}
{\cal H}_0&=&\int_{-\infty}^{\infty} \! \! \! dx \;
i v_F \psi_{\nu,\sigma}^\dagger(x) {\partial  \over \partial
x}\psi_{\nu,\sigma}(x) \; , \\
{\cal H}_\perp&=&\frac{1}{2} \left[ J_\perp^{(\nu)}
S^{+} \psi^\dagger_{\nu \downarrow}(0) \psi^{}_{\nu \uparrow}(0) + {\rm
H.c.} \right] \; , \\
{\cal H}_\parallel&=&\frac{1}{2} J_z^{(\nu)}
S_z \left[\psi^\dagger_{\nu \uparrow}(0) \psi^{}_{\nu \uparrow}(0) -
\psi^\dagger_{\nu \downarrow}(0) \psi^{}_{\nu \downarrow}(0) \right] \; ,
\\
{\cal H}_B&=&BS_z \! + \! \frac{B}{2} \! \! 
\int_{-\infty}^{\infty} \! \! \! \! \! \! \! dx \!
\left[\psi^\dagger_{\nu \uparrow}(x) \psi^{}_{\nu \uparrow}(x) -
\psi^\dagger_{\nu \downarrow}(x) \psi^{}_{\nu \downarrow}(x) \right]
\! . 
\end{eqnarray}
Summation is implicit over repeated channel indices $\nu=1,2$ and spin
configurations $\sigma=\uparrow,\downarrow$. $S^{\pm}$ and $S_z$
define the $S=1/2$ spinor at the origin. I have explicitly included a
coupling to an external magnetic field in ${\cal H}_B$ and allowed
asymmetry in both spin space [{\it e.g.} $J^{(\nu)}_\perp \ne J_z^{(\nu)}$]
and between channels [{\it e.g.} $J^{(1)}_\perp \ne J^{(2)}_\perp$]. 

I now proceed by bosonizing this model. This is a standard procedure
which I briefly review for chiral fermions moving
a ring of length $L$ with anti-periodic boundary
conditions $\psi(L/2)=-\psi(-L/2)$ (ensuring that $k_F=0$ defines
a unique ground state). Ultimately the limit $L \rightarrow \infty$
will be taken. 
As noted by Mattis and
Lieb\cite{mattis63} particle-hole excitations of a chiral 1D model may be
represented by bosonic variables
\begin{equation}
b^\dagger_q = \sqrt{2 \pi \over Lq} \sum_k \! : \!  c^{\dagger}_{k+q} c^{}_k
\! \! :, \quad 
b_q = \sqrt{2 \pi \over Lq} \sum_k \!  :\!  c^{\dagger}_k c^{}_{k+q}\!
\! : .
\end{equation}
These operators act on excitations above a vacuum state
where all states below $k_F$ are occupied and all those above are empty.
Normal ordering `::' with respect to this vacuum is essential 
to avoid the infinities associated
with the infinite number of occupied states. Here $q=2\pi n/L>0$. 
These bosons obey the correct
commutation relations $[b^{}_k,b^{\dagger}_q]=\delta^{}_{kq}$ by virtue of
the absence of a lower bound on the occupied momentum states. The
electron creation operators are defined such that
$\psi^\dagger(x)=L^{-1/2}\sum_k
e^{ikx} c^\dagger_k$. 

The idea behind bosonization is to use these bosons as an alternative
to the fermion basis for describing 1D models. The requirement that
this new basis be complete means that, in addition to the bosons which
describe particle-hole excitations, we also need 
ladder operators which create extra particles of a given spin and
flavor in the Fermi sea\cite{haldane81}, $F^\dagger_{i}$. 
These ladder operators (or ``Klein factors'') must also anticommute 
between different electron flavors
and spins to reflect the fermionic nature of the underlying particles
created. They may also be defined to be unitary. There are many
possible representations for the ladder operators, and I
chose to represent $F^\dagger_i$ by $(-1)^{\sum_{j<i}\hat{N}_j}
e^{i\hat{\theta}_i}$. Here the Hermitian
operator $\hat{\theta}_i$ is conjugate to the number operator $\hat{N}_i$:
$[\hat{\theta}_i,\hat{N}_j]=i\delta_{ij}$ and
$[\hat{\theta}_i,\hat{\theta}_j] = 0$. This formulation corresponds to
a particular ordering of the vacuum by the arbitrary assigning of an
order to the spin/flavor degree of freedom $i$ (e.g. $1,\uparrow$
comes before $2,\downarrow$).
By ensuring that the ladder operators commute with boson fields,
Haldane was able to show that one may represent a Fermi field as
\begin{equation}
\psi_i^\dagger(x) = \lim_{\alpha \rightarrow 0} {e^{ik_Fx} 
\over \sqrt{2\pi \alpha}} (-1)^{\sum_{j<i} \hat{N}_j} e^{i \Phi_i(x)} \; ,
\label{fermion}
\end{equation}
where
\begin{equation}
\Phi_i(x) = \hat{\theta}_i+{2 \pi x \hat{N}_i \over L}
+i \sum_{q>0} \! \sqrt{2 \pi \over Lq} \! \left[
e^{-iqx -q\alpha/2}b_{i,q} - {\rm H.c.} \right] \! .
\end{equation}
($\alpha$ is a short wavelength cut-off which regularizes the momentum
sums.) Together with Eq.~\ref{fermion}, the following relations form a
dictionary for bosonization
\begin{eqnarray}
\left[\Phi(x)_i,\Phi_j(y)\right]&=&-i\delta_{ij} \pi {\rm sgn}(x-y) \; , \\
\partial_x \Phi(x) =
\Phi'(x) &=& 2 \pi :\psi^\dagger(x)\psi^{}(x): \; , \\
\left[\Phi'(x),\Phi(y) \right] &=& -2 \pi i \delta(x-y) \; ,
\label{com1} \\
iv_F \int_{-L/2}^{L/2} \! \! \! \! dx \;
\psi^\dagger {\partial \over \partial x} \psi &=&
{v_F \over 4 \pi} \int_{-L/2}^{L/2} :\left[ \Phi'(x)\right]^2 : dx \; .
\end{eqnarray}

Using this dictionary, I now express the
Hamiltonian of Eq.~\ref{ham} in terms of phase fields for each species
of fermion. I choose $\{1\uparrow,2\downarrow,1\downarrow,2\uparrow\}$
as my ordering convention for the Klein factors. Any ordering will
work, but this one simplifies the computation as will be explained later.
Following both Emery and
Kivelson and Sengupta and Georges, the spin 
and charge degrees of freedom can be separated and both have symmetric
and antisymmetric combinations across the two channels:
\begin{eqnarray}
\Phi_{c_\pm}&=&{1\over 2}\left[ 
 \left( \Phi_{1\uparrow}+\Phi_{1\downarrow}\right)
\pm\left(\Phi_{2\uparrow}+\Phi_{2\downarrow}\right)
\right] \; ,  \nonumber \\
\Phi_{s_\pm}&=&{1\over 2}\left[ 
 \left( \Phi_{1\uparrow}-\Phi_{1\downarrow}\right)
\pm\left(\Phi_{2\uparrow}-\Phi_{2\downarrow}\right)
\right] \; .
\label{linear}
\end{eqnarray}
The number operators are defined analogously.

In terms of this decoupling, the Hamiltonian may be written as
\begin{equation}
{\cal H}_0={v_F \over 4\pi} \sum_{\lambda=\{c_\pm,s_\pm\}} 
\int_{-\infty}^{\infty} :\left[\Phi'_\lambda (x) \right]^2 : dx \; , 
\qquad \qquad
\label{ststart} 
\end{equation}
\vspace{-0.7cm}
\begin{eqnarray}
\lefteqn{{\cal H}_\perp = {J_\perp^{(1)} \over 4 \pi \alpha}
(-1)^{\hat{N}_{c_+} \! + \hat{N}_{s_-}} \! \! \left\{ \! S^+
e^{-i\left[\Phi_{s_+}(0)+\Phi_{s_-}(0)\right]} - {\rm H.c.} \right\}
{} }
\nonumber \\
& & {}
- \! {J_\perp^{(2)} \over 4 \pi \alpha}
(-1)^{\hat{N}_{c_-}\! \! -\hat{N}_{s_-}} \! \! \left\{ \! S^+
e^{-i\left[\Phi_{s_+}(0) -
\Phi_{s_-}(0)\right]} - {\rm H.c.} \right\}, 
\end{eqnarray}
\vspace{-0.7cm}
\begin{eqnarray}
{\cal H}_\parallel &=& {J_z^{(1)}\! +J_z^{(2)} \over 4\pi}
S_z \Phi'_{s_+}(0) +
{J_z^{(1)}\! -J_z^{(2)} \over 4\pi}
S_z\Phi'_{s_-}(0),  \\
{\cal H}_B&=&B\left[S_z + {1 \over 4\pi} \int_{-\infty}^{\infty}
\Phi'_{s_+}(x) dx \right] \; . \label{stend}
\end{eqnarray}
From this Hamiltonian, I will first derive the interacting resonant level
model obtained by Fabrizio {\it et. al.}\cite{fabrizio}. I will then
show the connection to the ($\sigma,\tau$) model.

The resonant level model is obtained by the unitary transformation
used by Emery and Kivelson
\begin{equation}
U = e^{-i \sigma_z \Phi_{s_+}(0)/2} \; ,
\end{equation}
where $\sigma_z$ is the 3rd Pauli spin matrix. The advantage of my original
choice of ordering is that, since $(-1)^{N_{s_+}}$ does not appear, this
transformation does not affect the Klein factors.

Using Eq.~\ref{com1}
and the fact that, if $[A,B]$ commutes with both $A$ and $B$, then
$[e^{\lambda A},B]=\lambda [A,B] e^{\lambda A}$, I find
\begin{equation}
U^\dagger \Phi'_{s+}(x)U=\Phi'_{s+}(x)-2\pi S_z \delta(x) \; .
\end{equation}
With ${\cal H'}=U^\dagger{\cal H}U$ I have, ${\cal H}_0'={\cal H}_0$
and 
\begin{eqnarray}
{\cal H}'_\perp &=& {J_\perp^{(1)} \over 4 \pi \alpha} 
(-1)^{\hat{N}_{c_+} + \hat{N}_{s_-}} \left[ S^+
e^{-i\Phi_{s_-}(0)} - {\rm H.c.}
\right] \nonumber \\
&&+ {J_\perp^{(2)} \over 4 \pi \alpha}
(-1)^{\hat{N}_{c_-}-\hat{N}_{s_-}} \left[ S^+
e^{i\Phi_{s_-}(0)} 
- {\rm H.c.} \right], \\
{\cal H'}_\parallel &=& \left( \bar{J}_z-2\pi v_F \right)
{S_z\over 2\pi}\Phi'_{s_+}(0) + \delta J_z {S_z \over
2\pi}\Phi'_{s_-}(0) \; , \\
{\cal H}'_B&=&{B \over 4\pi} \int_{-\infty}^{\infty}
\Phi'_{s_+}(x) dx  \; .
\end{eqnarray}
Here I have written $\bar{J}_z=(J_z^{(1)}+J_z^{(2)})/2$, $\delta
J_z=(J_z^{(1)} - J_z^{(2)})/2$ and will similarly define
$\bar{J}_\perp$ and $\delta J_\perp$. 

To obtain the equivalent resonant level model I `refermionize'
the problem by expressing the boson fields $\Phi_\lambda$ in terms of
fermion fields $\psi_\lambda$. For example, 
\begin{equation}
\begin{array}{cc}
\psi^\dagger_{s_-} = {e^{i\Phi_{s_-}(x)} \over \sqrt{2\pi\alpha}}
\; , &
\psi^\dagger_{s_+} = {(-1)^{\hat{N}_{s_-}} 
e^{i\Phi_{s_+}} \over \sqrt{2\pi\alpha}} \; . \\
\end{array}
\label{referm}
\end{equation}
(Again the ordering of these fermions is arbitrary.)
Notice that since there are
no terms in the Hamiltonian which change $N_{c_\pm}$, the operators
$(-1)^{\hat{N}_{c_\pm}}$ simply fix an arbitrary sign for
$J_\perp^{(1),(2)}$. Also note that for $S=1/2$, I can define
$S^+(-1)^{\hat{N}_{s_-}}= d^\dagger$, where $d^\dagger$ creates a
fermion which anticommutes with
$\psi^\dagger_{s_-}$. (Maintaining the Klein factors is crucial to
obtaining this relation.)

Thus finally I may write the equivalent 
resonant level model for the
two-channel anisotropic Kondo model as
\begin{eqnarray}
&{\cal H}&= iv_F \int_{-\infty}^{\infty} \left[ \sum_\lambda 
\psi^\dagger_\lambda
{\partial \over \partial x} \psi_\lambda+B
\psi^\dagger_{s_+}\psi^{}_{s_+} \right] dx \nonumber \\
&&+{\bar{J}_\perp \over 2i\sqrt{\pi \alpha}}\hat{b}
\left[\psi^\dagger_{s_-}+\psi_{s_-} \right]_{x=0}+{\delta J_\perp
\over 2\sqrt{\pi \alpha}}\hat{a} 
\left[\psi^\dagger_{s_-}-\psi_{s_-} \right]_{x=0}
\nonumber \\
&&+i\hat{a}\hat{b}\left[\left(\bar{J}_z -
2\pi v_F\right)\psi^\dagger_{s_+}\psi_{s_+}+\delta J_z
\psi^\dagger_{s_-}\psi_{s_-} \right]_{x=0} . \label{res}
\end{eqnarray}
The complex fermion $d$ has been expressed in terms of its real and
imaginary ({\it i.e.} Majorana)
components $\hat{a}=(d^\dagger+d)/\sqrt{2}$ and $\hat{b}
=(d^\dagger-d)/i\sqrt{2}$.
This emphasizes that the Majorana representation for the spin 
is perhaps the most
natural since the conduction sea is
seen to interact independently with the spin via its Majorana components.
This is the model obtained by Fabrizio {\it et. al.}\cite{fabrizio}
using the Anderson--Yuval technique---essentially identifying the
perturbation expansions of the two models~\cite{footy}. 
The charge degrees of freedom ($\lambda=c_\pm$) have completely
decoupled and remain free---an observation that will be exploited in
the second part of the paper. The Toulouse point corresponds to
$\bar{J}_z = 2\pi v_F$ and $\delta J_z=0$ where the model is non
interacting and may be solved exactly\cite{toulouse}. Here the impurity spin
completely decouples from the magnetic field so the local spin
susceptibility is zero. The constraint $\delta J_z=0$ means that in
the Toulouse point only describes the region in the vicinity of the
symmetric model.

This model was analyzed in full in Ref.~\onlinecite{fabrizio} for
the region around the Toulouse point. The two resonant level widths
$\Delta_b=\bar{J}_\perp^2/4\pi v_F \alpha$ and $\Delta_a=(\delta
J_\perp)^2/4\pi v_F \alpha$, associated with the two independent
Majorana modes, represent the two relevant energy scales of the problem.

\begin{figure}[btp]
\epsfxsize=3.325in \epsfbox{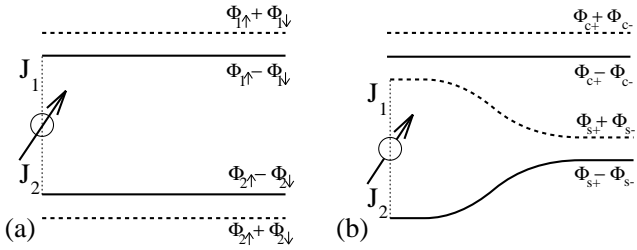}
\protect\caption{A pictorial representation of the relationship
between the two-channel Kondo model (a) and the $(\sigma,\tau)$ model
(b). By expressing the boson fields $\Phi$ in terms of spin and charge
degrees of freedom $(s_\pm,c_\pm)$ instead of the original flavor and
spin decoupling $(1_{\uparrow,\downarrow},2_{\uparrow,\downarrow})$ I
can refermionize the model to obtain the $(\sigma,\tau)$ formulation.}
\label{fig1}
\end{figure}

A second equivalent description of the problem
is obtained rather straightforwardly
from the Hamiltonian expressed in terms of spin and charge degrees of
freedom Eq.~\ref{ststart}--\ref{stend}. Refermionizing this
Hamiltonian using Eq.~\ref{referm}, one obtains
\begin{eqnarray}
&{\cal H}&=iv_F \sum_\lambda \int_{-\infty}^{\infty} \! \! \! \!
dx \psi^\dagger_\lambda
{\partial \over \partial x} \psi_\lambda + B\!\left[S_z+ {1\over 2} 
\int_{-\infty}^{\infty} \! \! \! \! \! \! \!
:\rho_{s_+}(x): dx \right] \nonumber \\
&&+{J_\perp^{(1)} \over 2} \left[S^+ \psi_{s_-} \psi_{s_+}
+ {\rm H.c.} \right] + {J_\perp^{(2)} \over 2} \left[S^+
\psi^\dagger_{s_-} \psi^{}_{s_+} + {\rm H.c.} \right]  \nonumber \\
&&+{J_z^{(1)} \over 2}
:\! \left[\rho_{s_+}(0)+\rho_{s_-}(0) \right] \! : + {J_z^{(2)} \over 2}
:\! \left[\rho_{s_+}(0)-\rho_{s_-}(0) \right] \! :  . \nonumber 
\end{eqnarray}
The normal ordering of the density operators is an instruction to
subtract off the contribution coming from the vacuum state. The mean
occupancy in the vacuum at $x=0$ is $1/2$ for any
fermion so I may replace $:\! \rho(x)\! :$ by $\rho(x)-1/2$. The
($\sigma$,$\tau$) description is apparent if I re-label subscripts $s$
and $c$ as channel indices 1 and 2, and subscripts $+$ and $-$ as
$\uparrow$ and $\downarrow$. This relabeling is shown pictorially in
Fig.~\ref{fig1}. I may then write the Hamiltonian, here
for the rotationally invariant case, $J_\perp=J_z$,  as 
\begin{eqnarray}
{\cal H}&=&{\cal H}_0 + \left[J^{(2)} \vec{\sigma}_1(0)+ J^{(1)}
\vec{\tau}_1(0) \right] \cdot \vec{S}  \nonumber \\
&&+\vec{B} \cdot \left[ \vec{S} + \int_{-\infty}^\infty
\vec{\sigma}_1(x)+\vec{\tau}_1(x) dx \right] \; ,
\end{eqnarray}
where
\begin{eqnarray}
\vec{\sigma_i}&=&
(\psi^{\dagger}_{i\uparrow},\ \psi^{\dagger}_{i\downarrow}) \cdot
{\vec{\sigma}\over 2} \cdot \left( \begin{array}{c} \psi^{}_{i\uparrow} \\
\psi^{}_{i\downarrow} \end{array} \right) \; ,\\
\vec\tau_i&=&
(\psi^{\dagger}_{i\uparrow}, \; \psi^{}_{i\downarrow}) \cdot
{\vec{\sigma}\over 2} \cdot \left( \begin{array}{c} \psi^{}_{i\uparrow} \\
\psi^{\dagger}_{i\downarrow} \end{array} \right) \; .
\end{eqnarray}

This is precisely the model treated in Ref.~\onlinecite{pedest} for the
symmetric case [$J^{(1)}=J^{(2)}$] and in Ref.~\onlinecite{schofield95}
in the general case [$J^{(1)}\ne J^{(2)}$]. The impurity is seen to
interact now with the spin and charge degrees of freedom of a single
chain leaving a second chain---representing what were previously charge
excitations---decoupled from the spin (see Fig.~\ref{fig1}).
Since both the
($\sigma,\tau$) description and the interacting resonant level models
can be derived from the original Kondo problem, they are both
equivalent representations of the model. Thus, for example, the low energy
spectrum of the $(\sigma,\tau)$ model plus the uncoupled chain will be
{\em identical} to that of the two channel Kondo model. 
What then are the relative merits of these two equivalent models over
the original formulation? 

By studying the non-interacting 
Toulouse point of the
resonant level description a qualitative picture of the thermodynamics
may be obtained. One sees a two-stage quenching of the entropy of the
impurity spin as the temperature is lowered through the two energy
scales. At temperatures between these two energy scales the
thermodynamic response is controlled by the fixed point of the channel
symmetric case. For example, the local spin susceptibility is
logarithmically divergent as a function of temperature. At low
temperatures the specific heat and susceptibility revert
to their Fermi liquid forms. These results mirror those obtained from
the exact solution. However, the disadvantage of this approach is that
to solve the model one is restricted to the region near
$J_z^{(1)}=J_z^{(2)} \gg J_\perp$ and so one may fail to identify properties
which are universal functions of the energy scales. For example,
the Wilson ratio is
vanishingly small near the Toulouse points in
contrast to the rotationally invariant case where the Wilson
ratio spans the range $(0,2]$.

It is in understanding the low temperature fixed point behavior that
the ($\sigma,\tau$) description excels. Unlike the
original formulation, the strong coupling fixed point of the ($\sigma,\tau$)
model is stable: over-screening of the impurity spin is
prevented by the Pauli principle. The intermediate coupling fixed
point of the two-channel Kondo model has been{\em moved} to infinity by the
transformation to the ($\sigma,\tau$) picture. One may then exploit a
strong coupling expansion to examine the physics near the fixed
point. Just as in the single channel case~\cite{nozieres,hewson}, symmetries
restrict the form of the resulting fixed point Hamiltonian. As well as
enabling a direct calculation of the Wilson ratio, one also obtains the
form of the low-energy Hamiltonian. The Majorana character of this
Hamiltonian is the basis for the conjecture that the excitations are
not compatible with weakly interacting fermion
quasi-particles~\cite{schofield95}. 

Finally I address the issue of how one uses the fixed point Hamiltonian
of the $(\sigma,\tau)$ model to compute the resistivity in the two-channel
Kondo model. The two models are equivalent as we have shown but
the mapping between them is linear only in the bosonized language of
Eq.~\ref{linear}. Put simply, the electrons in the two-channel Kondo
model are {\em not} the same electrons appearing in the $(\sigma,\tau)$
model.
To calculate the resistivity of the two-channel
Kondo model 
one must show how the electron Green's function transforms under this mapping. 
It does not transform into the electron Green's function in the 
$(\sigma,\tau)$ model as implicitly assumed in Ref.~\onlinecite{zhang96}. 
The appropriate correlator is expressible only in terms
of the boson fields: for example
\begin{eqnarray}
{\cal G}_{1,\uparrow}(x_1,x_2;\tau)={\cal T}_\tau 
\langle \psi_{1,\uparrow}(x_2,\tau)\psi^\dagger_{1,\uparrow}(x_1,0)\rangle
, \quad \quad \nonumber \\
\rightarrow
\sim {1 \over 2\pi\alpha} {\cal T}_\tau \langle :\! e^{-
i\sum_\lambda \Phi_\lambda (x_2,\tau)/2}:
 :\! e^{i\sum_\lambda \Phi_\lambda (x_1,0)/2}: \rangle \; ,
\end{eqnarray}
where $\lambda=\{c_\pm,s_\pm\}$. Following Affleck and 
Ludwig~\cite{affleck93}, the low temperature 
resistivity in the isotropic two-channel Kondo model is found 
by considering this
correlator with $x_2 < 0 < x_1$ in the fixed point Hamiltonian 
of Ref.~\onlinecite{pedest}. The leading
temperature dependence comes from consideration of the leading irrelevant
operator which affects this correlator at {\em first order} and, since it
has dimension $1/2$, leads to a $T^{1/2}$ correction~\cite{schofield96}.

In conclusion, two methods have been proposed recently to give a
simple account of the physics of the two-channel Kondo model both at
and away from the channel symmetric point. These methods sidestep, at
the expense of completeness, the full formalism of the Bethe Ansatz by
considering models related to the original two-channel Kondo problem.
In this paper these related models are formally derived from the
original two-channel Kondo model via bosonization. In describing the
bosonization technique, care is taken to include explicitly the Klein
factors which preserve the anticommutation relations between the
fermions on different channels. As expected, the interacting resonant
level model found previously by Fabrizio {\it et. al.} is found as a
simple generalization of the Emery-Kivelson Hamiltonian. In addition,
however, the ($\sigma,\tau$) model with its strong coupling fixed
point is also obtained directly from the two-channel Kondo model by
bosonization. This establishes the equivalence of this model to the
original two channel Kondo model. The relative merits of these two
approaches are discussed.

During the course of this work I have benefitted enormously from
discussions with N. Andrei, P. Azaria, P. Coleman, A. Jerez,
G. Kotliar, Ph. Nozi\`eres, A. Sengupta and Q. Si. 
I also gratefully acknowledge
the financial support of a
Royal Society (NATO) travel fellowship and NSF grants 
DMR93--12138 and DMR92--21907. 

\end{document}